\documentclass[reprint,aps,pre]{revtex4-2}

\usepackage{dcolumn}
\usepackage{bm}

\usepackage[utf8]{inputenc}
\usepackage[T1]{fontenc}
\usepackage{mathptmx}
\usepackage{etoolbox}
\usepackage{amsmath}

\usepackage{graphicx}    
\usepackage{siunitx}
\usepackage{xcolor}
\usepackage{float} 
\usepackage{tabularx}
\usepackage{amssymb}
\usepackage{ulem}
\usepackage{multirow}

\usepackage[version=3]{mhchem} 
\usepackage{chemformula} 

\usepackage{subcaption}
\newcommand{\corr}[1]{#1}

\begin{document}

\title{Fluorescent molecular rotor-based polymer materials for local microviscosity mapping in microfluidic channels}
  
\author{Dharshana Nalatamby}
\author{Florence Gibouin}
\author{Maxime Zitouni}
\author{Julien Renaudeau}
\author{Gérald Clisson}
\author{Pierre Lidon}
\affiliation{Laboratoire du Futur, (LOF) - Syensqo - CNRS - Université de Bordeaux, UMR 5258, Bordeaux, 33600 Pessac, France.}
\author{Simon Harrisson}
\email{sharrisson@enscbp.fr}
\affiliation{Univ. Bordeaux, CNRS, Bordeaux INP, LCPO, UMR 5629, Bordeaux, 33600 Pessac, France}
\author{Yaocihuatl Medina-Gonzalez}
\email{yaocihuatl.medina-gonzalez@u-bordeaux.fr}
\affiliation{Laboratoire du Futur, (LOF) - Syensqo - CNRS - Université de Bordeaux, UMR 5258, Bordeaux, 33600 Pessac, France.}

\date{\today}

\begin{abstract}

A viscosity-sensitive monomer consisting of a methacrylate-functionalized julolidine-based molecular rotor (MECVJ) was synthesized and used to obtain viscosity-sensitive polymers (poly(DMA-\textit{s}-MECVJ)). The qualitative properties of the molecular rotor were preserved after its inclusion in the new polymer, in particular the effect of the viscosity of the surrounding medium on the fluorescence lifetime of the rotor. By grafting these polymers onto glass slides, viscosity-sensitive surfaces were obtained, showing good robustness in time after successive use and washing. As proof of concept, these surfaces were used to assemble a microfluidic chip capable of mapping local microviscosity of fluids flowing inside the channel. 
 
\end{abstract}

\maketitle

\section{Introduction}
\label{intro}
Viscosity is an intrinsic property of fluids, which determines the behavior of flows~\cite{Pathikonda2021} and whose measurement can give important information about molecular composition and small-scale interactions. Consequently, viscosity measurement is essential in domains as varied as polymer synthesis and processing, crude oil processing, ink jet and 3D printing, cosmetic product formulation, slurries for battery fabrication and food processing~\cite{Meffan2023, Dietrich2010, Fan2022,  Li2022}. For example, in the medical domain, changes in viscosity may indicate the presence of pathological conditions~\cite{Meffan2023, Wu2023}. Mechanical rheometers or viscometers can be used to measure viscosity at the macroscale, but measurements at the microscale require specialized microviscometers that are prone to artifacts and some systems cannot be characterized with these tools. Other measurement techniques have been developed, for example  electrochemical and electrokinetic microviscometers, but they may damage samples that are sensitive to heating or to the application of an electrical potential. Other techniques, designed for microfluidic and lab-on-a-chip applications, are based on the measurement of flow velocity under an imposed flow rate, and include capillary-driven flow microviscometers with vibrational~\cite{NguyenTV2015}, thermal~\cite{Puchades2011},  rotational~\cite{Kuenzi2011} or optical detectors~\cite{Bamshad2018}. However, all of these approaches are not fully satisfactory to assess local variations of viscosity, as they all rely on averaged measurements over some volume, possibly small yet macroscopic.

Alternatively, microrheology techniques are based on the analysis of the motion of micro- or nanoscale probe particles suspended in the medium, under brownian motion are forces exerted by external fields. By assuming Stokes-Einstein equation to relate particle displacement to rheological properties of the surrounding material, this motion can be related to the viscous and elastic properties of the medium. In general, these techniques have limitations concerning the resolution of spatial heterogeneities or applications to stiff and non-equilibrium systems, and they probe mechanical properties at an intermediate length scale corresponding to the probe size~\cite{Mao2022}.

Fluorescent molecular rotors (FMR) have recently been used as viscosity probes in systems that are not accessible to mechanical viscometers, including aerosols~\cite{Kuimova2013}, microfluidic channels~\cite{Nalatamby2023,Haidekker2016}, confined fluids~\cite{Florence2024} and living cells~\cite{Kuimova2012}, enabling to map viscosity locally with excellent spatial resolution. After excitation by light, FMRs  can undergo conventional radiative fluorescent decay, or non-radiative relaxation; the latter mode is accompanied by a twist and internal rotation of the molecule. Interactions with the surrounding fluid hinder molecular rotation, favoring relaxation through the radiative path and increasing the observed fluorescence intensity and lifetime. As a result, these molecules are sensitive probes of local viscosity. After calibration of the fluorescence intensity or lifetime in fluids of known viscosity, FMRs can be used to determine the viscosity of unknown samples, and map spatial and temporal variations in the viscosity of complex systems~\cite{Nalatamby2023, Haidekker2016, Strickler1962, Haidekker2007, Haidekker2010}.
FMR can thus be considered as one additional technique that can afford information in systems where other microheology techniques present limitations~\cite{Xia2018,Ciccuta2007,Waigh2016,Furst}.
FMR have for instance been used in other studies concerning the monitoring and understanding of chemical reactions in particular, monomers sensitive to viscosity have been synthesized and used as probes in polymerizations, using fluorescence imaging techniques such as FLIM~\cite{Lopez2024}.

To date, FMR have primarily been used in solution, with only a single report of the viscosity-responsive properties of 9-(2-carboxy-2-cyanovinyl)julolidine (CCVJ), a widely used FMR, immobilized on optical fibers~\cite{haidekker2006optical}. Grafting of FMR on polymers would open new opportunities to design viscosity sensors: such an approach has already been used in Kwak’s group, yet with no attempt to attach the polymer to surfaces~\cite{Lee_Kwak2011,Jin_Kwak2017,Jin_Kwak2019}.
This raises the question of the sensitivity of FMR in complex environments. In molecular fluids, FMR are expected to sense the mechanics of their cybotactic environment, which is directly correlated with macroscopic viscosity. In complex, non-newtonian fluids, for which macroscopic viscosity results from interactions at supramolecular scale, some studies tackled this question by characterizing response of free FMR in polymer solutions~\cite{Bittermann2021}, during polymerization~\cite{Nolle2014}, or upon approaching glass transition~\cite{Mirzahossein2022}, or of FMR grafted on glass surfaces under nanometric confinement~\cite{Suhina2015,SuhinaWeber2015}.
In particular Suhina et al synthesized rigidochromic molecules sensitive to viscosity that were then covalently attached to glass cover slips. These probe-functionalized glass covers were used to visualize the contact of PMMA beads with these surfaces. They suggested that response of rotor is related to microscopic correlation lengths (confinement size, blob size in polymer systems, etc.) and thus indirectly to the macroscopic viscosity of the sample.

The objective of the work presented here is to fabricate a viscosity-sensitive surface by grafting it with an FMR-functional polymer. The obtained viscosity-sensitive materials were incorporated into a microfluidic device to allow the measurement and mapping of the viscosity of fluids flowing through its channels. This strategy paves the way to a new generation of simple and affordable viscosity sensors for applications concerning microfluidics.

\section{Materials and Methods}

All chemicals used were obtained from Sigma-Aldrich and used as received. 

\subsection{Synthesis of the viscosity-sensitive monomer (MECVJ) \textbf{(4)}}

\label{subsec1}
A monomer, 9-2-(methacryloxyloxy)ethyloxycarbonyl-2-cyanovinyl-julolidine (MECVJ), was synthesized via condensation of an aldehyde and a methacrylate ester, which were themselves synthesized as described in the following paragraphs.

\subsubsection{Synthesis of Julolidine Aldehyde 
\textbf{(2)}}
\label{subsubsec1}

\begin{figure*}[!htb]
\centering
\includegraphics[width=\textwidth]{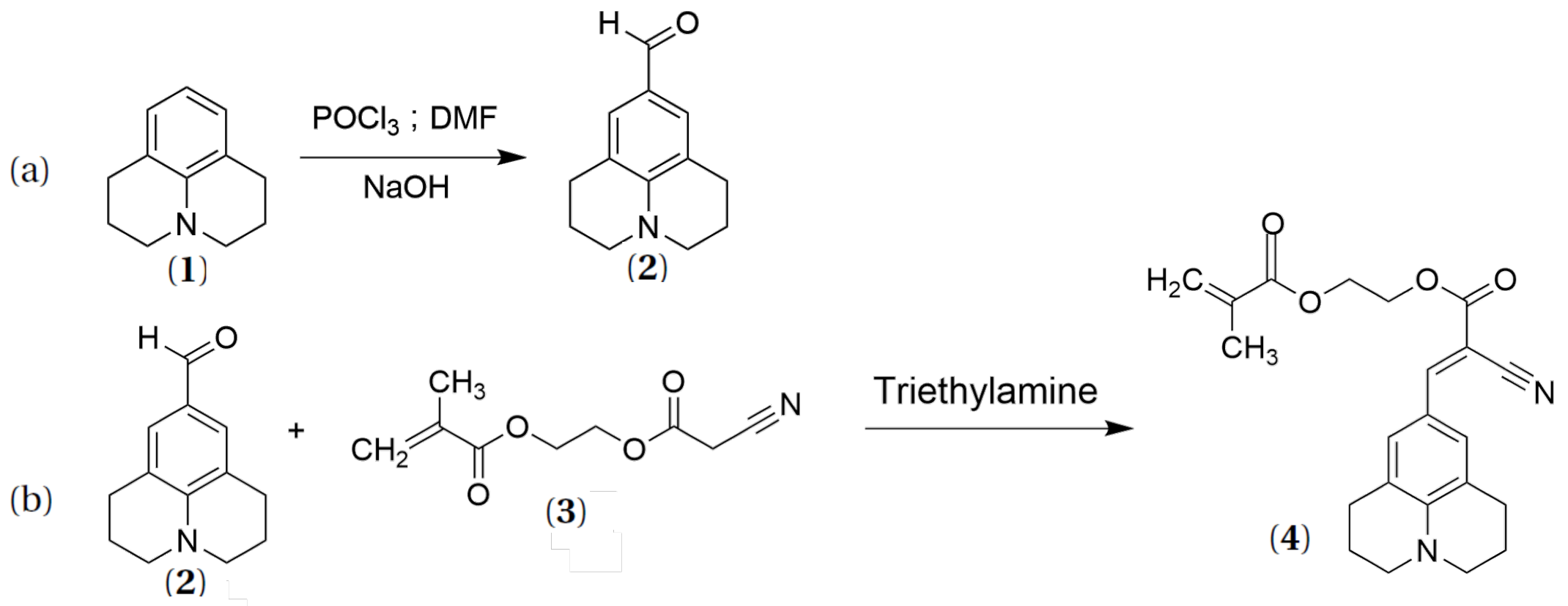}
\caption{(a) Synthesis route of the julolidine aldehyde(\textbf{2}) from julolidine (\textbf{1}). (b) Synthesis route of MECVJ (\textbf{4}) via the condensation of julolidine aldehyde (\textbf{2}) and methacrylate ester (\textbf{3}).}\label{synthesis}
\end{figure*}

The protocol for the synthesis of the julolidine aldehyde was adapted from the work of Haidekker et al~\cite{Haidekker2001} (see Schema~\ref{synthesis}(a)). Julolidine ($\SI{0.5}{\gram}$, $\SI{2.87}{\milli\mole}$) was placed in a mixture formed by $\SI{0.54}{\gram}$ ($\SI{7.4}{\milli\mole}$) of N,N-dimethylformamide (DMF) and $\SI{7}{\milli\liter}$ ($\SI{0.11}{\mole}$) of dichloromethane (DCM). Phosphorous oxychloride (POCl$_{3}$, $\SI{0.58}{\milli\liter}$ ($\SI{2.4}{\milli\mole})$) was added dropwise. The mixture was stirred for $\SI{12}{\hour}$ at $\SI{25}{\celsius}$. The flask was then placed in an ice bath for $\SI{15}{\min}$ before adding $\SI{28}{\milli\liter}$ of an aqueous solution of sodium hydroxide (\ch{NaOH}, 2 M). The mixture was stirred at $\SI{0}{\celsius}$ for $\SI{4}{\hour}$.

\textsuperscript{1}H-NMR was used to monitor the reaction advancement and to characterize the final product. These analyses were performed  in a Bruker Avance III HD 400 spectrometer with a field frequency of $\SI{400.2}{\mega\hertz}$. For this purpose, samples were dissolved in CDCl\textsubscript{3} at a concentration of about $\SIrange{10}{12}{\milli\gram\per\milli\liter}$. Tetramethylsilane was used as internal standard at 0 ppm. Figure S1 
shows the NMR spectra of julolidine and of the aldehyde. The characteristic peak for \ch{-CH=} band of julolidine can be observed at $\SIrange{6.53}{6.57}{ppm}$ (t, \textsuperscript{1}H) and the characteristic peak of aldehyde appears at $\SI{9.53}{ppm}$ (s, \textsuperscript{1}H), which confirms the absence of julolidine in the final product. The purity of julolidine aldehyde was estimated at $\SI{97}{\percent}$. The peak at $\SI{5.52}{ppm}$ corresponds to traces of DCM used as solvent, which did not cause any problem for the next synthesis. 

\subsubsection{Synthesis of the Methacrylate Ester \textbf{(3)}}

The synthesis of the ester was performed through a modification of the work by Tsakos et al~\cite{Tsakos2015}, who used 1-ethyl-3-(3-dimethylaminopropyl)carbodiimide hydrochloride (EDC) to induce the esterification. For that, $\SI{40}{\milli\liter}$ ($\SI{0.8}{\mole}$) of acetonitrile and $\SI{2.9}{\gram}$ ($\SI{24}{\milli\mole}$) of hydroxyethylmethacrylate (HEMA) were added to $\SI{2.8}{\gram}$ ($\SI{42}{\milli\mole}$) of cyanoacetic acid and $\SI{6.3}{\gram}$ ($\SI{33}{\milli\mole}$) of EDC. This reaction mixture was kept at $\SI{40}{\celsius}$ and stirred during $\SI{91}{\hour}$, then acetonitrile was removed in a rotary evaporator. The obtained solid was dissolved in DCM and washed twice with $\SI{50}{\milli\liter}$  of HCl; then twice with $\SI{60}{\milli\liter}$ of water, twice with $\SI{60}{\milli\liter}$ of saturated sodium bicarbonate solution and once with $\SI{60}{\milli\liter}$ of saturated sodium chloride solution. The organic layer was obtained and dried with \ch{MgSO}\textsubscript{4}. \textsuperscript{1}H-NMR was used to verify the absence of by-products and of residual agents. DCM was evaporated under reduced pressure and the methacrylate ester was obtained as pale yellow oil. The yield of this reaction was $\SI{77}{\percent}$ and the purity of the final product was around $\SI{98}{\percent}$. 

\subsubsection{Synthesis of the viscosity-sensitive monomer (MECVJ) \textbf{(4)}}

The viscosity-sensitive fluorescent monomer, MECVJ was synthesized by the route shown in Schema~\ref{synthesis}(b). To this purpose, $\SI{0.51}{\gram}$ ($\SI{2.57}{\milli\mole}$) of 2-cyanoacetoxyethylmethacrylate and $\SI{0.59}{\gram}$ ($\SI{2.95}{\milli\mole}$) of julolidine aldehyde were added to $\SI{15}{\milli\liter}$ ($\SI{185}{\milli\mole}$) of tetrahydrofuran. The mixture was kept at $\SI{50}{\celsius}$ and $\SI{1.6}{\milli\liter}$ ($\SI{11.5}{\milli\mole}$)  of triethylamine were added drop-wise. Stirring was continued for $\SI{96}{\hour}$. The solvent was then evaporated at reduced pressure. The residue was recrystallized twice from a mixture of DCM/heptane (1:10), yielding a bright orange solid. The yield of this reaction was around $\SI{42}{\percent}$ and the approximate purity of more than $\SI{95}{\percent}$. \textsuperscript{1}H-NMR spectrum is shown in Figure S2. 

\subsection{Synthesis and characterization of viscosity-sensitive fluorescent polymers poly(DMA-\textit{s}-MECVJ)}

\subsubsection{Synthesis}

Four different polymers of different target molecular weights were synthesized, labeled from P1 to P4 as shown in Table~\ref{table:summary-polymer}. The initiator used was azobisisobutyronitrile (AIBN). The required amounts of monomers, MECVJ and DMA (dimethylacrylamide), initiator [I] and RAFT (Reversible Addition Fragmentation Chain Transfer) agent were calculated to obtain the desired ratio of equivalent numbers as [MECVJ]/[DMA]/[I]/[RAFT]. For instance, for P3, $\SI{31.9}{\milli\gram}$ ($\SI{0.10}{\milli\mole}$) of RAFT agent and $\SI{3.8}{\milli\gram}$ ($\SI{0.023}{\milli\mole}$) of AIBN were added to $\SI{9}{\milli\liter}$ ($\SI{106}{\milli\mole)}$ of dioxane in a twin-neck round bottom flask. Then, $\SI{994}{\milli\gram}$ ($\SI{10.03}{\milli\mole)}$ of DMA and $\SI{38.9}{\milli\gram}$ ($\SI{0.137}{\milli\mole)}$ of MECVJ were added to the flask. The reaction medium was then degassed by bubbling with N\textsubscript{2} during $\SI{30}{\min}$. The flask was heated to $\SI{60}{\celsius}$ and stirred to start the polymerization reaction. To monitor the conversion of MECVJ and DMA, samples for \textsuperscript{1}H-NMR analyses were obtained via cannula transfer under a current of N\textsubscript{2}. Quenching of the polymerization was obtained by immersion of the flask in an ice bath. Removal of dioxane under reduced pressure afforded a dark yellow viscous liquid, which was solubilized in $\SI{10}{\milli\liter}$ of DCM and added drop-wise to $\SI{200}{\milli\liter}$ of diethyl ether. Precipitation of the polymer was then observed. Centrifugation at $\SI{2500}{rpm}$ and at $\SI{5}{\celsius}$ during $\SI{30}{\min}$ allowed to obtain a dark orange solid. 

\begin{table*}[!htb]
\begin{center}
\begin{tabular}{|c|c|c|c|c|c|c|c|} 
 \hline
   &$n_{\text{targeted}}$ &\centering DMA/MECVJ ($\SI{}{\mole\percent}_\text{initial}$) & $t \, (\SI{}{\hour})$ & Conv.DMA ($\SI{}{\percent}$) & $M_{n}^\text{[SEC]} \, (\SI{}{\gram\per\mole})$ & $M_{n}^\text{[theor]} \, (\SI{}{\gram\per\mole})$ &  \DJ \\ 
 \hline
 P1 & 20 & \centering 100/1 & 27 & 93 & 4575 & 2219 & 1.19 \\ 
 \hline
 P2 & 50 & \centering 95/5 &  16 &  85 & 3697 & 5270 & 1.21  \\
 \hline
 P3 & 100 & \centering 100/1 &  6 & 81 & 9240 & 8645 & 1.18  \\ 
 \hline
P4 & 100 & \centering 90/10 &  54 & 81 & 4742 & 11346 & 1.76 \\
 \hline
\end{tabular}
\caption{Details of the four poly(DMA-\textit{s}-MECVJ) statistical copolymers synthesized at $\SI{60}{\celsius}$ and their macromolecular characteristics. The total number of monomers (MECVJ and DMA) targeted in a polymer chain is denoted as $n_{\text{targeted}}$,  $t$ is the polymerization reaction time in hours. The conversion of DMA ($\SI{}{\percent}$) was determined from \textsuperscript{1}H-NMR. Masses $M_{n}^\text{[SEC]}$ and $M_{n}^\text{[theor]}$ are polystyrene equivalents, determined by SEC in DMF/LiBr, calibrated using polystyrene standards. \DJ \, is the polydispersity index.}
\label{table:summary-polymer}
\end{center}
\end{table*}

\subsubsection{Characterization of the poly(DMA-\textit{s}-MECVJ)}

\paragraph{NMR} 
\textsuperscript{1}H-NMR spectroscopic analyses were performed with a Bruker Avance III HD 400 spectrometer with a field frequency of 400.2 MHz. The analyzed polymers were dissolved in approximately $\SI{0.5}{\milli\liter}$ of \ch{CDCl3}, and the concentration of the polymer solutions was 10-12 $\SI{}{\milli\gram\per\milli\liter}$. The samples were referenced to TMS (0 ppm).

\paragraph{Size Exclusion Chromatography (SEC)} Polymer molar masses were determined by SEC using N,N-dimethylformamide (DMF + LiBr $\SI{1}{\gram\liter}$) on an Ultimate 3000 system from ThermoFischer Scientific (Ilkirch, France) equipped with a differential refractive index detector from Wyatt technology (Santa Barbara CA, USA). Polymers were separated on three Shodex Asahipack gel columns [GF 310 ($7.5 \times \SI{300}{\milli\meter}$), GF510 ($7.5 \times \SI{300}{\milli\meter}$), exclusion limits from $\SIrange{500}{300000}{Da}$] at a flowrate of $\SI{0.5}{\milli\liter\per\minute}$. The column temperature was held at $\SI{50}{\celsius}$. An Easivial™ kit of Polystyrene from Agilent (Santa Clara CA, USA) was used as calibration standard ($M_n$ from $162$ to $\SI{364000}{Da}$) to provide polystyrene-equivalent molecular weights.

\subsection{Characterization of response to viscosity}

In order to quantify the response of MECVJ and poly(DMA-\textit{s}-MECVJ) to the viscosity of the surrounding medium, solutions of these FMR were prepared in DMSO/glycerol mixtures. For that, stock solutions of MECVJ and of poly(DMA-\textit{s}-MECVJ) in DMSO were prepared beforehand, and dispersed in DMSO-glycerol mixtures. Care was taken to keep a similar concentration of FMR in all solutions, equal to $\SI{e-5}{\mole\per\liter}$, in order to avoid bias on fluorescence intensity due to changes of concentration.

\subsection{Viscosity measurements}

In order to determine viscosity, ramps of shear rate were applied to the solutions by using a Kinexus Ultra+ rheometer (Netzsch). Samples of low viscosity (below $\SI{15}{\milli\pascal\second}$) were analyzed using a double-gap geometry whereas cone-plate geometry was used for viscous samples (above $\SI{15}{\milli\pascal\second}$). A Peltier module was used to regulate the experimental temperature, which was fixed at $\SI{25}{\celsius}$.

Both for solutions of MECVJ and of polymer poly(DMA-\textit{s}-MECVJ) in mixtures of glycerol and DMSO, viscosity was found to be independent on shear rate. This Newtonian behavior for polymer solution is not surprising as the added polymer is strongly diluted. The average over all of the applied shear rates was taken as the final viscosity value. 

\subsubsection{Photophysical characterization}

For the different solutions, absorption and emission spectra were measured using a Cary UV-Vis spectrophotometer and a Cary Eclipse Fluorescence spectrophotometer both from Agilent Technologies. Fluorescence intensities $I_\text{F}$ were measured as the maximum value in emission spectra. The intensities reported in this study are not corrected by measured absorption and could thus be biased by fluctuations of FMR concentration from sample to sample. To avoid this, great care has been taken to keep this concentration identical and no evidence of such bias has been observed.

Lifetimes $\tau_\text{F}$ were measured using a LIFA system for Fluorescence Lifetime Imaging Microscopy (FLIM) from Lambert Instruments, equipped with a LED source of wavelength $\SI{451}{\nano\meter}$, close to the maximum of absorption of the FMR. This instrument allows the mapping of fluorescence lifetimes across the field of view of a microscope, the lowest detection limit of this equipment is \SI{1}{\nano\second}.
The lifetimes $\tau_\text{F}$ of the samples were measured by recording a reference beforehand; a fluorescein aqueous solution ($\SI{10}{\micro\mole\per\liter}$, Sigma-Aldrich 46955-1G-F) prepared in $\mathrm{p}H$ 10 buffer solution with tabulated lifetime $\tau_\text{ref} = \SI{4.02}{\nano\second}$. FLIM acquisitions were carried out using a modulation frequency of $f = \SI{40}{\mega\hertz}$ and $12$ acquisition phases with a $1 \times$ CCD gain, ensuring maximum resolution.

\subsubsection{Förster-Hoffmann model}

The relationship between fluorescence intensity $I_\text{F}$ and lifetime $\tau_\text{F}$ of FMR, and viscosity $\eta$ is commonly described by a power-law relation~\cite{Nalatamby2023}. Briefly, the relationship between the quantum yield ($\phi_\text{F}$) of a molecular rotor and the viscosity of the surrounding medium can be simplified into an expression called the Förster-Hoffman equation~\cite{Loutfy1986,Haidekker2010}:
\begin{equation}
  \log \phi_\text{F} = \alpha \log {\eta}+ C
  \label{eq:FH-equation}
\end{equation}
\noindent in which $\alpha$ is a constant usually dependent on the rotor and $C$ is a proportionality constant. As $I_\text{F}$ and $\tau_\text{F}$ are proportional to $\phi_\text{F}$, the Förster-Hoffman expression can then be rewritten as :
\begin{equation}
  \log I_\text{F} = \alpha \log {\eta}+ C_{I_\text{F}}
  \label{eq:FH-Intensity}
\end{equation}
\noindent and :
\begin{equation}
 \log \tau_\text{F} = \alpha \log {\eta}+ C_{\tau_\text{F}} .
  \label{eq:FH-Lifetime}
\end{equation}

While coefficients $C_{I_\text{F}}$ and $C_{\tau_\text{F}}$ are difficult to interpret and depend on instrumental factors, the coefficient $\alpha$, which corresponds to the exponent of the power-law, is expected to be similar for both parameters $I_\text{F}$ and $\tau_\text{F}$ and quantifies the sensitivity of the FMR to the surrounding viscosity. Equations~\eqref{eq:FH-Intensity} and~\eqref{eq:FH-Lifetime} were thus used to analyze the response of FMR.

Equation~\eqref{eq:FH-Lifetime} is particularly valuable for applications, as $\tau_\text{F}$ is independent of the local concentration of the rotor and of other instrument-related factors such as local light intensity. It can thus be used for viscosity measurements of unknown samples provided a prior calibration has been established using similar fluids of known viscosities.

\subsection{Grafting of poly(DMA-\textit{s}-MECVJ) materials to glass surfaces and its use in microfluidics devices fabrication}

A "grafting-to" strategy based on the work by Duwez et al~\cite{Duwez2006} was adopted for the grafting of the viscosity-sensitive polymers, poly(DMA-\textit{s}-MECVJ, to glass surfaces. For that, a thin film of gold of $\SI{30}{\nano\meter}$ was deposited on a glass slide by metal evaporation with a pressure of $\SI{e-6}{\milli\bar}$ in a glove box. A chromium film of $\SI{5}{\nano\meter}$ of thickness was prior deposited onto the glass slide to ensure a good adhesion of the gold. A solution of polymer in water ($\SI{14.3}{\milli\gram\per\liter}$) was deposited onto the gold-coated glass slide, and left in a closed Petri dish at room temperature for at least $\SI{16}{\hour}$. This step was carried out in the dark to prevent possible photobleaching. The glass slide was then rinsed with $\SI{10}{\milli\liter}$ of distilled water and dried with N\textsubscript{2}.

To qualitatively confirm the grafting of poly(DMA-\textit{s}-MECVJ), contact angle measurements were performed using the sessile drop method using the Attension Theta Optical Tensiometer, as changes of hydrophilicity would confirm the presence of polymers on the glass slide.

To further analyze grafting efficiency, X-Ray Photoelectron Spectroscopy (XPS) analyses were performed by using a VG ESCALAB 220i-XL X-ray photoelectron spectrometer with non-monochromatic MgK$\alpha$ radiation ($h \nu = \SI{1253.6}{\electronvolt}$) and a pass energy of $\SI{80}{\electronvolt}$. An internal reference of C1s binding energy ($\SI{285.0}{\electronvolt}$) was used. The lateral resolution of the setup was of $\SI{150}{\micro\meter}$ in routine in ultra-high vacuum conditions at $\SI{e-8}{\milli\bar}$. These XPS analyses were performed on surfaces grafted from a solution of $\SI{5}{\milli\gram\per\liter}$ of polymer in water. 
Fluorescence properties of the prepared surface were measured with the FLIM-LIFA. 

\begin{figure}[!htb]
\centering
\includegraphics[width=0.5\textwidth]{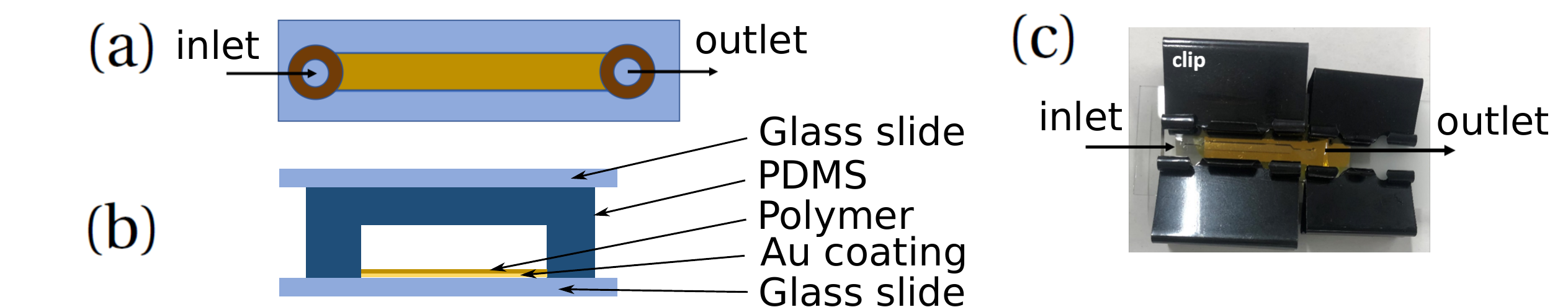}
\caption{Schematics of the viscosity-sensitive microfluidic chip fabricated (not at scale), (a) top view, (b) side view. (c) Photograph of the viscosity-sensitive PDMS microfluidic chip. Length of the field of view is about $\SI{8}{\centi\meter}$.}
\label{fig:schemachip}
\end{figure}

A viscosity-sensitive microfluidic chip prototype was then fabricated, as schematized in Figure~\ref{fig:schemachip}(a,b). Using standard soft lithography techniques,  a straight microfluidic channel of width $\SI{2}{\milli\meter}$, height $\SI{100}{\micro\meter}$ and length $\SI{2}{\centi\meter}$ was patterned in a PDMS block. This block was then sealed by clamping it with a glass slide grafted with poly(DMA-\textit{s}-MECVJ). The response of the surface was calibrated with DMSO/glycerol solutions of known viscosity and Förster-Hoffman model was applied to the obtained responses. Then, as a proof of concept, the response obtained by using the microfluidic channel was studied for different fluids as follows. Two different water/glycerol solutions of different viscosities, respectively of $\SI{2}{\milli\pascal\second}$ and $\SI{843}{\milli\pascal\second}$, were alternatively injected with a flow rate $Q \sim \SI{10}{\micro\liter\per\minute}$, with time period long enough to ensure the injected fluid flushed the previous one (corresponding to flushing about 40 times the volume of the main channel).

\section{Results and Discussion}

\subsection{Photophysical properties of the viscosity-sensitive monomer MECVJ}

\begin{figure*}[!htb]
\centering
\includegraphics[height=5cm]{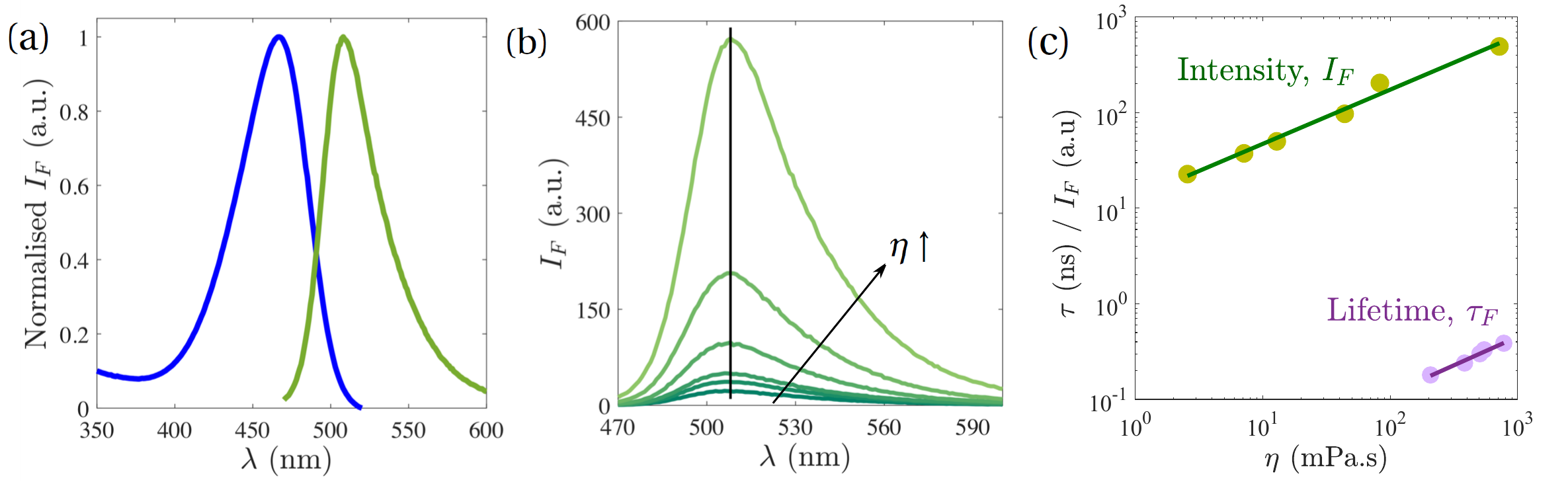}
\caption{(a) Excitation and emission spectra of MECVJ in glycerol. (b) Emission spectra of MECVJ in DMSO-glycerol mixtures (of viscosities $\eta = 3$, $7$, $46$, $44$, $43$ and $\SI{718}{\milli\pascal\second}$). The black solid line was added to indicate the unchanged maxima of the emission spectra. (c) MECVJ fluorescence intensity ($I_\text{F}$) and lifetime ($\tau_\text{F}$) as function of the viscosity ($\eta$) of the surrounding media. Solid lines are respectively power law fits along Eq.~\eqref{eq:FH-Intensity} and~\eqref{eq:FH-Lifetime}\corr{, $\alpha=0.6$}.}
\label{fig:caracMECVJ}
\end{figure*}

The absorption and emission maxima of the monomer MECVJ were found at $\SI{467}{\nano\meter}$ and $\SI{508}{\nano\meter}$ respectively as shown in Figure~\ref{fig:caracMECVJ}(a), similar to what was reported for julolidine-based FMR, meaning that the fluorescence properties were not affected by the addition of the methacryloyloxyethyl group to the (cyano-vinyl) julolidine base. The polymers were tested in a variety of DMSO-glycerol mixtures, of varying viscosities without changes in the maximum wavelengths of absorption and emission (Figure~\ref{fig:caracMECVJ}(b)) while the maximum intensity was observed to $I_\text{F}$ increase with viscosity.

Figure~\ref{fig:caracMECVJ}(c) shows the fluorescence intensity $I_\text{F}$ and fluorescence lifetime $\tau_\text{F}$ of MECVJ as function of viscosity. The Förster-Hoffman model was found to be valid in the range of viscosity used for calibration. The coefficient obtained ($\alpha=0.6$) was similar to those reported for other julolidine-based FMR, such as DCVJ in glycerol/ethylene glycol mixtures of different viscosities ~\cite{Kung1986,Haidekker2000}; values obtained for $C$ parameter were $C_{I_\text{F}}=1.1$ and $C_{\tau_\text{F}}$ = -2.1 when $\eta$ is expressed in $\SI{}{\milli\pascal\second}$ and $\tau$ in $\SI{}{\nano\second}$. 

\subsection{Synthesis and characterization of the viscosity-sensitive copolymer poly(DMA-\textit{s}-MECVJ)}

\subsubsection{Synthesis of poly(DMA-\textit{s}-MECVJ)}

The viscosity-sensitive copolymers were synthesized using RAFT (reversible addition-fragmentation chain transfer) polymerization~\cite{Moad2013}. They contained \numrange{1}{10}\% of MECVJ and \numrange{90}{99}\% of dimethyl acrylamide (DMA), and had targeted degrees of polymerization ranging from 20 to 100. DMA was chosen as its polymer is soluble in a wide range of polar and nonpolar solvents, ensuring that the copolymerized MECVJ units would be accessible to the solvents that are to be analyzed. An additional benefit of using an acrylamide comonomer is that acrylate and acrylamide monomers are consumed less rapidly than methacrylates in copolymerization~\cite{Ercole2010, Trang2007}. This ensures near-quantitative conversion of the MECVJ even when full conversion of the DMA is not reached, reducing the risk of contamination of the copolymer with unreacted MECVJ.

\begin{figure}[!htb]
\centering
\includegraphics[scale=0.3]{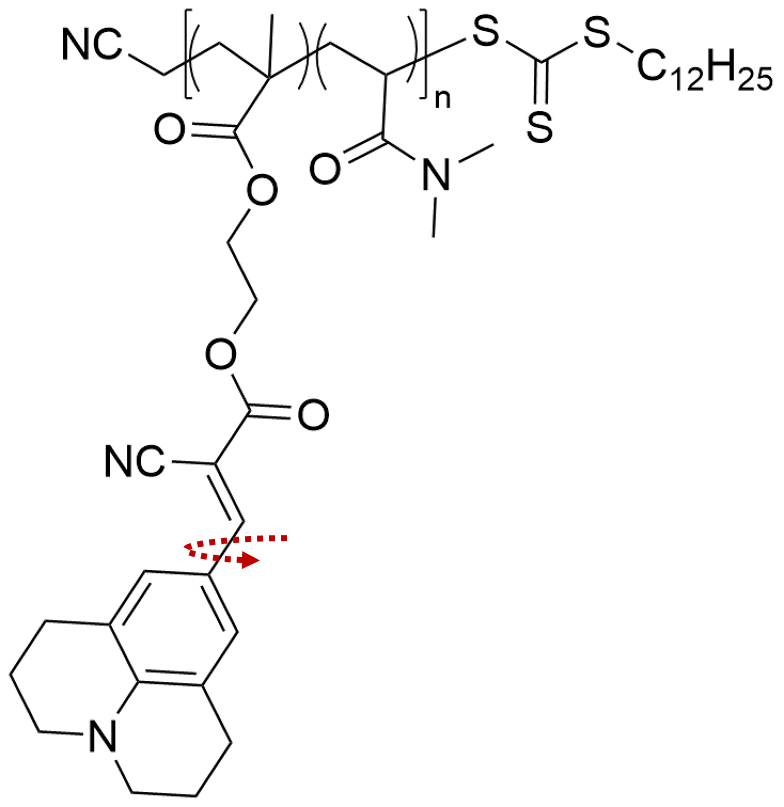}
\caption{General structure of the targeted viscosity-sensitive polymer, poly(DMA-\textit{s}-MECVJ). The dotted arrow shows the intended rotation of the julolidine part.}
\label{genstructurepolym}
\end{figure}

Monitoring of the polymerization kinetics was performed using $^1\text{H-NMR}$ and results are displayed in Supporting Information (see Fig. S3) for polymer P2. Full conversion of DMA is reached after about 13 hours of reaction. The quality of the retrieved polymers was further analyzed with $^1\text{H-NMR}$ and SEC, and the results obtained for P3 are provided in Supporting Information (see Fig. S4). An absence of starting monomers and solvent can be perceived, and the narrow differential refractive index signal in the chromatograms confirms the low dispersity of the synthesized polymer. Nevertheless, as listed in Table~\ref{table:summary-polymer}, a low dispersity value was not achieved for P4, for which MECVJ makes up around $\SI{10}{\percent}$ of its composition. Cyanomethyl dodecyl trithiocarbonate is a mediocre RAFT agent for methacrylate monomers, it could be that there was too much MECVJ in this polymerization, and thus, control was lost. However, it would be more likely that the fluorescent part of the MECVJ interacts with the propagating radicals in some way. Lower than expected molecular weights and higher than expected dispersities may indicate irreversible chain transfer reactions, while retardation may indicate increased chain termination. This reasoning could explain the lower than expected molar masses $M_n$ obtained for P2 and P4. 

At the end of the synthesis, the polymers obtained after precipitation contained no detectable unreacted MECVJ or DMA. SEC analyses indicated that some control over molecular weight was obtained, as polymers with a higher ratio of RAFT agent to monomer displayed lower molecular weights. Reasonable agreement between the experimental molecular weight and the theoretical molecular weight, calculated from the conversion of each monomer, as well as relatively narrow molecular weight distributions (\DJ$\approx1.2$) were observed for P2 and P3. Higher than expected molecular weight, combined with low dispersity, was obtained for the polymer with the lowest targeted molecular weight (P1), which may indicate some fractionation during precipitation leading to loss of the lowest molecular weight fraction of the polymer. The polymer with the highest concentration of MECVJ (P4) exhibited a broad molecular weight distribution and lower than expected $M_n$, suggesting that significant termination occurred during this polymerization. The loss of control of the polymerization may be due to interactions between MECVJ and the propagating radicals, leading to irreversible chain transfer or other chain-terminating reactions.

\subsubsection{Photophysical properties of poly(DMA-\textit{s}-MECVJ)}

\begin{figure*}[!htb]
\centering
\includegraphics[height=5cm]{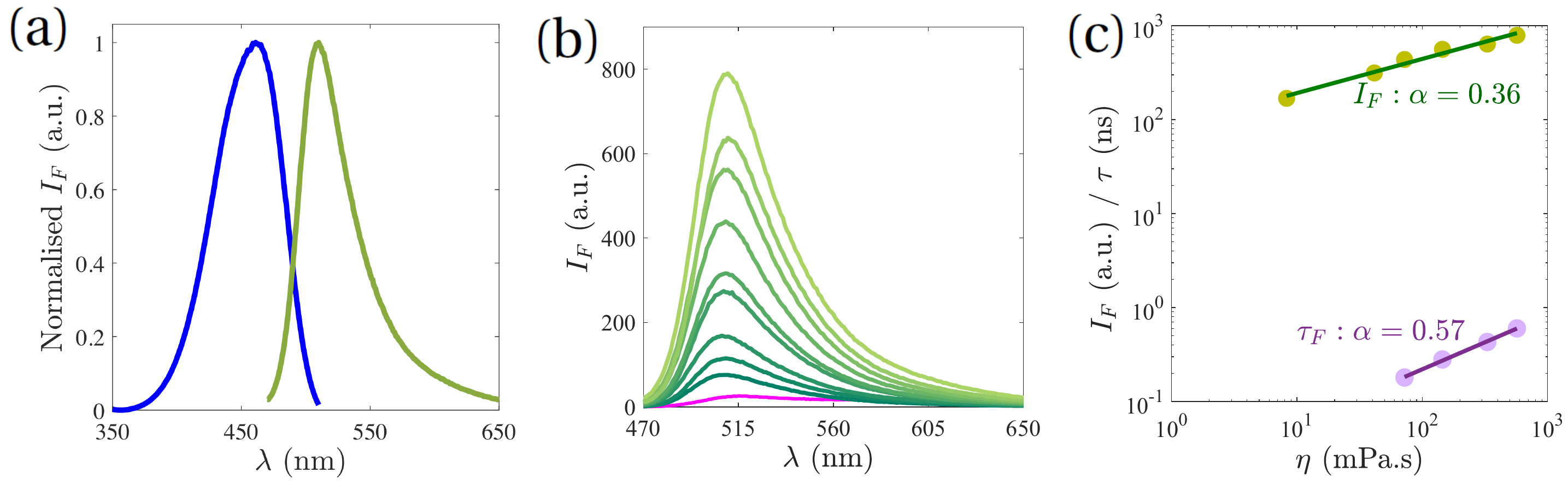}
\caption{(a) Absorption (blue line) and emission (green line) spectra of P2 in glycerol. Maximum of absorption and emission are respectively $\SI{461}{\nano\meter}$ and $\SI{510}{\nano\meter}$. (b) Emission spectra of P2 in different glycerol/DMSO mixtures of respective viscosity $\eta = 1$, $2$, $5$, $8$, $13$, $41$, $72$, $145$, $330$ and $\SI{569}{\milli\pascal\second}$ with excitation wavelength $\SI{450}{\nano\meter}$. (c) Corresponding calibration curves in logarithmic scale of P2. Intensity $I_\text{F}$ is represented with green points and lifetime $\tau_\text{F}$ with purple points.}
\label{fig:carac_P2}
\end{figure*}

The absorption and emission maxima were obtained for the different synthesized polymers (see Figure~\ref{fig:carac_P2}(a) for polymer P2 and Table S1 in Supporting Information) and found to be similar to the wavelengths obtained for MECVJ in solution, respectively around $\lambda_\text{abs} = \SI{460}{\nano\meter}$ and $\lambda_\text{em} = \SI{505}{\nano\meter}$.

Figure~\corr{\ref{fig:carac_P2}(b)} shows the variations obtained in fluorescence emission spectra of \corr{polymer P2} as function of the viscosity of the surrounding medium \corr{(results for other polymers can be found in Figures S5,6,7(a) of SI)}. The obtained results were in overall agreement with these obtained for the MECVJ \corr{and qualitatively similar for the different polymers}, which confirmed that the fluorescence properties are maintained after polymerization. The length of the polymer chain did thus not hinder the response of the polymer and the wavelength of maximum emission of the polymers was independent of the viscosity of the samples. 

As for MECVJ, it was observed that the emission wavelength is similar in the different solutions, which confirmed that only viscosity influenced the quantum yield $\phi_\text{F}$ of these polymers and hence their lifetime $\tau_\text{F}$ in the solutions used. \corr{It is to note a peculiarity of polymer P4, for which a shoulder develops around $\lambda \sim \SI{580}{\nano\meter}$ in the emission spectrum for solutions of high viscosity. As this polymer has the highest rotor density along the chain, this could be related to a coupling between them: this point would deserve further photophysical investigation, yet it does not compromise the sensitivity of fluorescence response to viscosity.}

Double logarithmic plots of fluorescence intensity and lifetime as function of viscosity are shown in Figure~\corr{\ref{fig:carac_P2}(c) for polymer P2 (see Figures S5,6,7(b) in SI for other polymers)}. The Förster-Hoffman relationship was found to be valid for the polymers synthesized in the range of viscosities tested. More precisely, comparable yet smaller $\alpha$ coefficients were found for polymers in comparison with free MECVJ, indicating lower sensitivity of the polymers to viscosity. 

At high concentration of MECVJ, for polymers P2 and P4, a discrepancy was observed between the coefficients $\alpha$ measured for intensity and for lifetime. This could arise from a non-exponential decay of the polymers fluorescence, reflected differently in the lifetime and in the intensity measurements~\cite{Lutjen2018}. \corr{Such a hypothesis in not easy to test directly with the frequency-domain FLIM used in this study, as lifetime is measured by analyzing phase delay between a modulated excitation light source and fluorescence response. Polar plot representation has been suggested as a means to evidence multi-exponential decay~\cite{lakowicz_1990}. For the polymers under study, the result indeed hints at behavior, but the interpretation remains ambiguous and deeper photophysical investigation would be required. If this was confirmed, the effective lifetime obtained by frequency-domain FLIM would be a nonlinear average of the physical relaxation rates. However, this does not compromise the conclusions of this study as results shown in Figures 5 and S5 to S7 show a clear correlation between this effective parameter and the local viscosity, and the synthesized polymers can still be used for viscosity sensing.}

Although the polymers were slightly less sensitive to viscosity than the monomer, they could still be proposed for quantitative measurement purposes and allowed to prepare viscosity-sensitive surfaces. Moreover, the lifetime values measured for the polymers were \corr{$\approx \SI{4}{\nano\second}$} longer than those obtained for MECVJ; this phenomenon arises from the rigidity of MECVJ when it is integrated in the poly(DMA-\textit{s}-MECVJ), which favors the emissive relaxation pathway, increasing as well the time that the FMR stays in the excited state and consequently $\tau_\text{F}$. \corr{Restriction of the motion of rotors inside polymers may be at the origin of the lower sensitivity of the MECVJ when embedded in them, this behavior has been observed before by M. Kubankova et al ~\cite{Kubankova2019} when BODIPY-based rotors were covalently linked to the HaLo protein. This phenomenon may be  circumvented by better engineering the polymers to allow good swelling in the presence of the solvent and better contact of the rotors with the fluid.}

This can be an advantage in the case of low viscosity solutions, associated with lifetime below the resolution limit of the FLIM. For instance, given the resolution limit of the LIFA equipment (limited to lifetime $\tau_\text{F} > \SI{1}{\nano\second}$), MECVJ could be used to measure viscosities larger than $\SI{200}{\milli\pascal\second}$, while the use of the polymers allowed to test viscosities ten times lower. 

\subsection{Grafting of poly(DMA-\textit{s}-MECVJ)}

\subsubsection{Quality of grafting and sensitivity to viscosity}

After grafting the polymer onto the gold-coated glass surfaces, contact angles decreased from around $\SI{70}{\degree}$ to around $\SI{30}{\degree}$, as the polymer increased the hydrophilicity of the surface. \corr{For the sake of clarity, XPS spectra for a surface obtained by grafting P3 are displayed in Fig. S4 of Supporting Information}. The increase of the atomic percentage of carbon and oxygen, the decrease in the atomic percentage of gold and the appearance of chemisorbed sulfur at $\SI{162.4}{\electronvolt}$ were attributed to the presence of grafted polymer~\cite{Fustin2009}. No leaching of the polymer was observed after thoroughly washing the surface with water. 

\corr{After grafting, polymer P4 afforded the best surface, in terms of homogeneity and of fluorescence response. Consequently, despite the unexplained appearance of a shoulder in its emission spectrum at large viscosity, this polymer was selected to fabricate the viscosity-sensitive surfaces and microfluidic chips presented in the following results.}

\begin{figure*}[!htb]
\centering
\includegraphics[height=5cm]{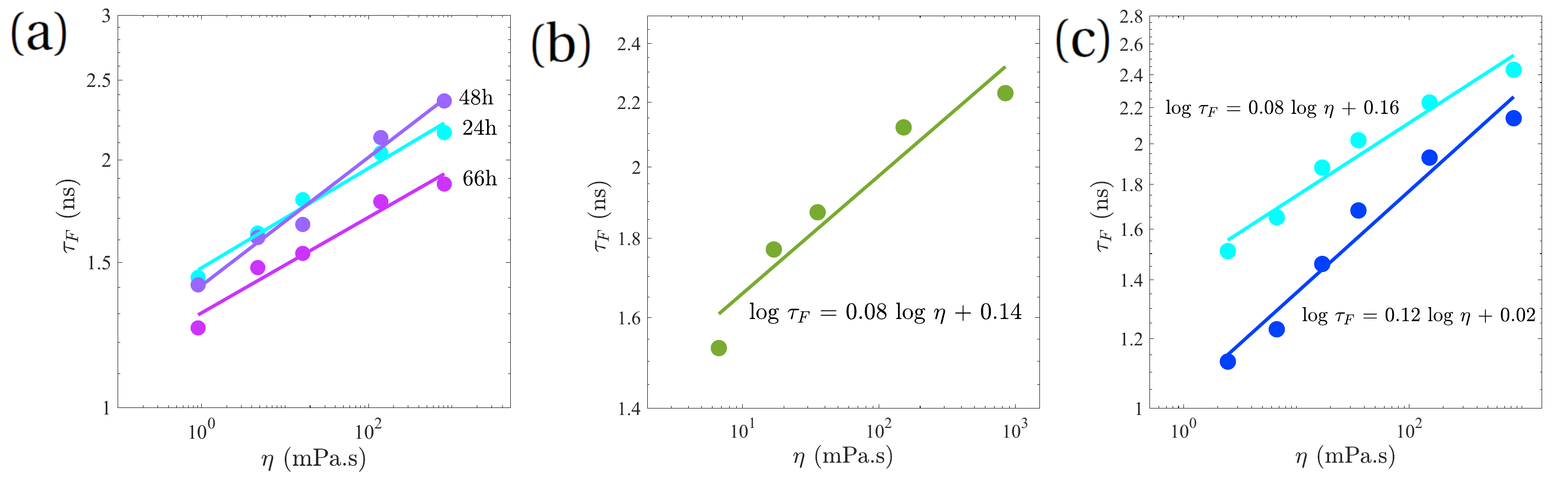}
\caption{\corr{(a) Fluorescence lifetime response of grafted viscosity-sensitive polymer to mixtures of different viscosity for various grafting times. Solid lines correspond to fit along Eq.~\eqref{eq:FH-Lifetime}. (b) Fluorescence lifetime response of grafted viscosity-sensitive polymer to mixtures of different viscosity after continuous washing for 63 hours.  Data are plotted in logarithmic scale with $\eta$ expressed in $\SI{}{\milli\pascal\second}$ and $\tau_\text{F}$ in $\SI{}{\nano\second}$. The straight line represents the Förster-Hoffmann model fit of Eq.~\eqref{eq:FH-Lifetime}. (c) Calibration curves in logarithmic scale of fluorescence lifetimes, $\tau_\text{F}$ in two different poly(DMA-\textit{s}-MECVJ)-grafted microfluidic chips versus water-glycerol mixture viscosity, $\eta$. The straight lines correspond to fits with the Förster-Hoffmann model with similar coefficient $\alpha \approx 0.1$ for both microdevices when $\eta$ is expressed in $\SI{}{\milli\pascal\second}$ and $\tau_\text{F}$ in $\SI{}{\nano\second}$.}\label{fig:FH_surfaces}}
\end{figure*}

Drops of various water/glycerol mixtures of different viscosities were deposited onto the surface obtained after grafting of P4. The excitation wavelength was set at $\SI{451}{\nano\meter}$ in the FLIM-LIFA and fluorescence lifetimes for each solution were recorded. The results obtained are shown in Figure~\corr{\ref{fig:FH_surfaces}(a)}. Figure~\corr{\ref{fig:FH_surfaces}(b)} shows the response of lifetime $\tau_\text{F}$ to mixtures of different viscosities for different grafting times. After continuous washing for $\SI{63}{\hour}$, grafted surfaces still shown satisfactory response to viscosity of flowing solutions (Figure~\corr{\ref{fig:FH_surfaces}(b)}), indicating that the polymer was robustly grafted. Coefficient $\alpha$ for grafted polymers was decreased compared to polymer in solution, reaching values around $0.1$. Despite this loss of sensitivity to viscosity after grafting, it remained sufficient to fabricate viscosity-sensitive surfaces and consider to design a microfluidic chip capable of measuring the viscosity of a fluid flowing through it. 

\corr{Bittermann et al have used non-attached, free molecular rotors to investigate the viscosity of polymer solutions and observed that an increase in polymer chain length may lead to a higher confinement at the nanoscopic scale. These authors proposed that the rotors were contained in “blobs” of polymers and the suggested confinement would have an impact on the rotors response. This interpretation may not be applicable in our case, as the rotors are not free but attached to the polymer chains and these polymer chains are attached to a surface, in this case, “blobs” formation is maybe not possible.}

\subsubsection{Towards a viscosity-sensitive microfluidic chip}

As a proof of concept, the obtained surface with grafted polymers \corr{P4} was used to prepare \corr{microfluidic devices as represented on Figure~\ref{fig:schemachip}(c). Channels molded in PDMS were clamped over the surface, and connected to syringe pumps to control fluid flow, with flowrate $Q \sim \SI{10}{\micro\liter\per\minute}$.}

Figure~\corr{\ref{fig:FH_surfaces}(c)} shows the variation of lifetime $\tau_\text{F}$ measured for different mixtures of water-glycerol of varying viscosity using the FMR-integrated microfluidic chip developed in this work. In order to check the repeatability of the process, the experiments were repeated with two chips prepared following the same protocol. The viscosity-sensitivity of the polymeric surface was thus not modified by its integration into the microfluidic device.

\begin{figure*}[!htb]
\centering
\includegraphics[height=5cm]{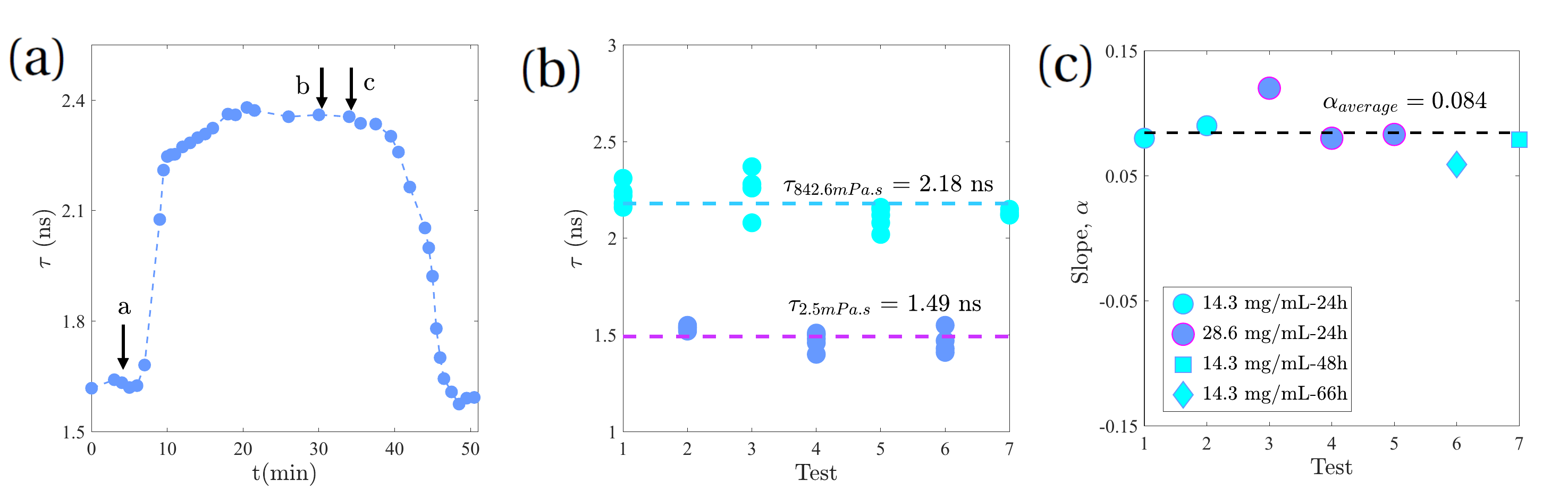}
\caption{\corr{(a) Real-time fluorescence response \sout{of viscosity-sensitive microfluidic chip developed in this thesis} to changes in viscosity of the microfluidic chip developed in this study; a further explanation on points (a,b, and c) is provided in the accompanying text. (b) A plot of fluorescence lifetime values obtained for two different mixtures of glycerol-water tested multiple times on the poly(DMA-\textit{s}-MECVJ)-grafted surface. The average lifetime values for both mixtures are represented by the dotted lines. (c) Comparison of coefficient-$\alpha$ of the Förster-Hoffmann model (slope-value) observed for polymer-grafted surfaces obtained with solutions of polymer at different concentrations and at different times of grafting. The dotted line represents the average value of $\alpha$ for all the represented data.}}
\label{fig:chip_carac}
\end{figure*}

The dynamic response of the developed microfluidic device to viscosity was further investigated, as depicted in Figure~\corr{\ref{fig:chip_carac}(a)}. The response of the polymer-grafted microfluidic chip during the injection of the $\eta \approx \SI{2}{\milli\pascal\second}$ (lower-viscosity) solution was measured before the injection of the $\SI{843}{\milli\pascal\second}$ (high-viscosity) solution at time (a). Time (b) marks the end of the injection of high-viscosity fluid, and at time (c), the lower-viscosity fluid was reintroduced into the microchannel. For both fluid injections, after a delay related to the time required by the fluid to reach the observation area, variations of lifetime $\tau_\text{F}$ were observed, consistently with the expected viscosity changes.

The response of the chip to the viscosity of the different fluids injected shows the satisfactory response of the polymer to the viscosity of the fluids flowing through the chip. Figure~\corr{\ref{fig:chip_carac}(b)} shows the successive lifetimes $\tau_\text{F}$ obtained at a fixed position in the microchannel by switching back and forth the injection between two different water/glycerol mixtures.

It is worth highlighting that the values of $\tau_\text{F}$ obtained for the tested mixtures did not change throughout seven successive tests, with the total duration lasting almost 3 hours, showing the good stability of the microfluidic device. A video obtained during the injection of the different solutions is presented as Supporting Information. 
\corr{It is to note that the response of the grafted rotor to the viscosity of the surrounding medium may be influenced by several parameters as interactions between the surface and the rotors, guided by, for instance, surface charges, the degree of swelling of the polymer and thus the entanglement of the polymer chains and the degree of exposition of the rotor to the flowing flow; as well, covalent linking of the rotor to the polymer may decrease the intramolecular twisting ability of the grafted rotor and directly reduce the viscosity effect on the quantum yield. Other phenomenon that can be present is that the rotors grafted onto the polymers may have access only to the boundary layer and not to the bulk of the fluid.
The  $\alpha$ coefficient value, determined during calibration, is directly related to the sensitivity of the material to the viscosity of the flowing fluid and needs to be cautiously determined. To avoid problems arising from a decrease of sensitivity, polymers may be judiciously chosen and engineered to expose all the rotors to the flowing fluids.}

Values of the $\alpha$ coefficient of the Förster-Hoffman relationship were obtained for the different tests performed with different polymer-grafted chips and had a mean value of $0.084$, as shown in Figure~\corr{\ref{fig:chip_carac}(c)}. The values of $\alpha$ for various viscosity-sensitive surfaces, grafted with polymer solutions of different concentrations and several grafting durations confirm the reproducibility of the results and stable sensitivity to viscosities. The values for $\alpha$ obtained in this work are comparable with and also more reproducible than those obtained in previous studies~\cite{Lichlyter2009}. As well, the polymeric material reported in this work may exhibit better sensitivity to viscosity than the materials published previously even after grafting onto glass surfaces. It is to note that Kwak's group obtained molecular-rotors grafted polymers with good sensitivity to viscosity and developed polymeric films with proved response to viscosity, however no grafting onto surfaces was performed by this group~\cite{Lee_Kwak2011, Jin_Kwak2017, Jin_Kwak2019}. 

\section{Conclusions and Perspectives} 

Viscosity-sensitive FMRs were successfully integrated into polymeric materials, namely poly(DMA-\textit{s}-MECVJ) by previously synthesizing a viscosity-sensitive monomer called MECVJ and incorporating it during the polymerization. The obtained materials, monomer and polymer, kept the sensitivity to viscosity from the julolidine group.  These polymeric materials were successfully anchored onto glass surfaces previously coated with a thin gold film. The surfaces thus obtained were sensitive to the viscosity of fluids flowing through a microchannel fabricated with the polymer-anchored surfaces.

While the polymeric surface presented less sensitivity to the viscosity of the surrounding fluid than the monomer alone, these materials were still useful to fabricate robust viscosity-sensitive surfaces and be used in a micro-channel to obtain a measurement device capable to map the viscosity of a fluid flowing inside the microchannel. 

\corr{Even if the resistance of these materials has not been optimized, in this work it has been observed that 1) after 63h of washing with water, the polymers presented sensitivity to viscosity and their response followed the Foster-Hoffmann equation, 2) after 50 min of tests, the prepared microfluidic chip exhibited no changes in the lifetime measured and 3) after 3 hours of continuous utilization by switching solutions, no photobleaching of the rotors was observed.}

This work represents a step towards the fabrication of viscosity-sensitive polymeric materials where fluorescence lifetime and intensity are related to the viscosity of the fluid in contact. Other polymers could be considered to give flexibility to the polymer chain and maintain the sensitivity of the rotor, or even to optimize the anchoring. \corr{The strategy presented here paves the way to a new generation of simple and affordable viscosity sensors for applications concerning microfluidics such as in lab on chip medical, diagnosis among others. }

\acknowledgements{The research presented in this article was funded by the ANR grant MicroVISCOTOR (ANR-18-CE42-0010-01) and by the MIELS project funded by the Nouvelle Aquitaine Region (project n° 205086). PLACAMAT facility is acknowledged for XPS analysis. Pr. J-C. Baret and Dr. J. Yuan from the CRPP and Dr. B. Cabannes-Boué from the LCPO are thanked for the metallization of glass slides. The authors also thank Solvay and CNRS for funding.}

\newpage

\appendix

\subsection{Synthesis of the Julolidine Aldehyde}

\textsuperscript{1}H-NMR spectra of julolidine (\textbf{1}) and of julolidine aldehyde (\textbf{2}).

\begin{figure}[!htb]
    \centering
    \includegraphics[width=0.5\textwidth]{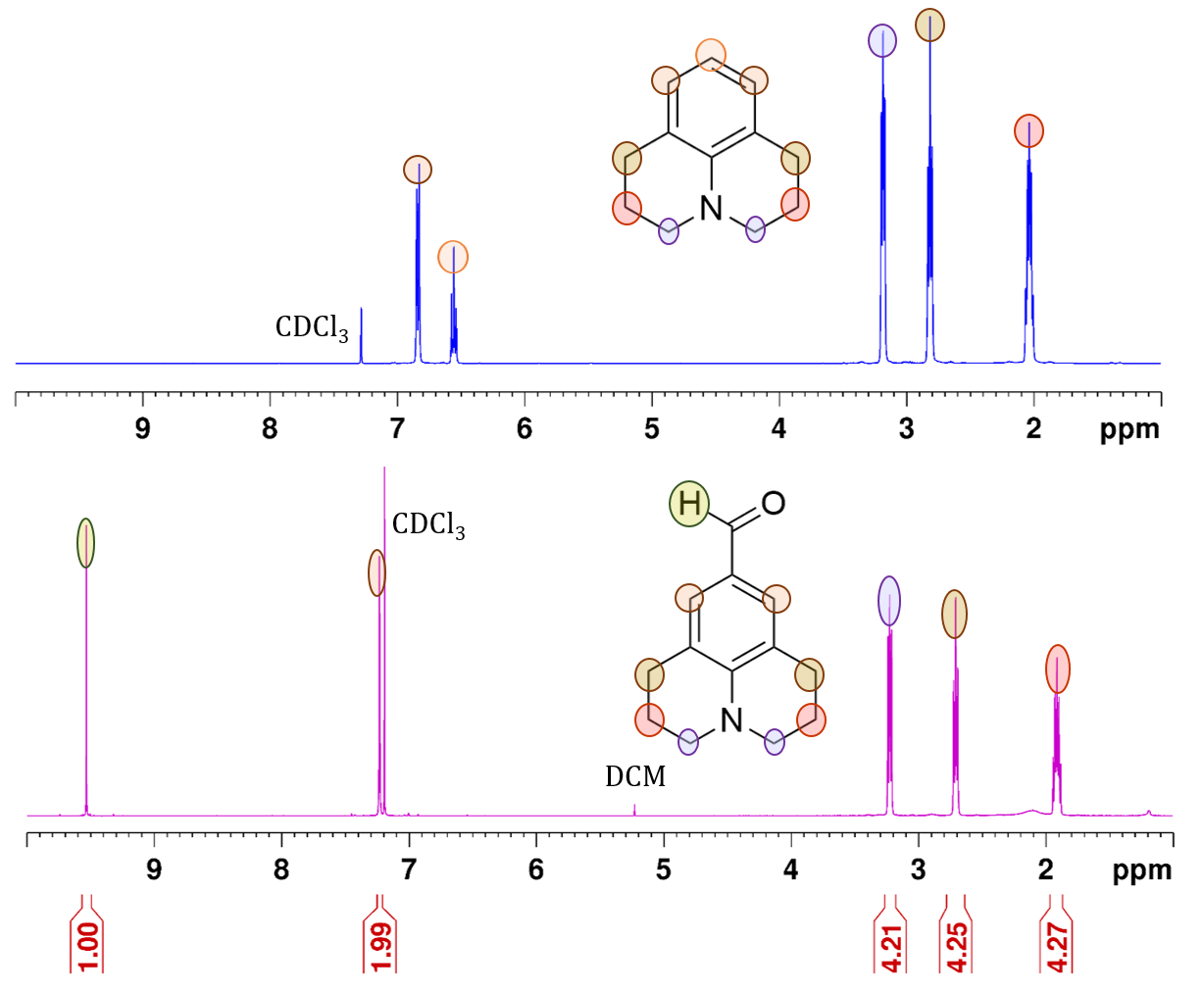}
    \caption{\textsuperscript{1}H-NMR spectra of julolidine (\textbf{1}) and of julolidine aldehyde (\textbf{2}) }
    \label{juloandaldehyde}
\end{figure}

\subsection{Synthesis of the viscosity-sensitive monomer (MECVJ)}

\textsuperscript{1}H-NMR spectra of the synthesized viscosity-sensitive fluorescent monomer MECVJ.

\begin{figure}[!htb]
    \centering
    \includegraphics[width=0.5\textwidth]{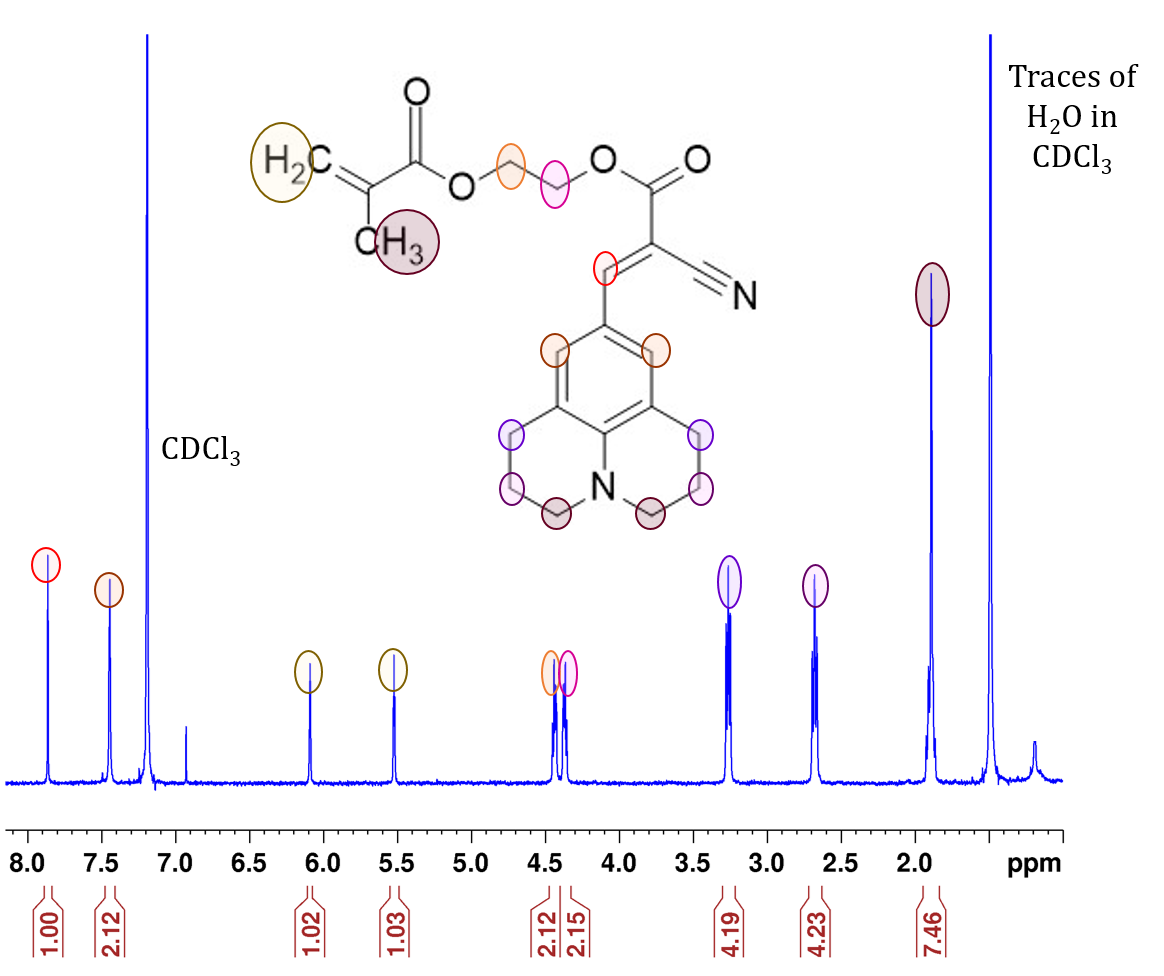}
    \caption{\textsuperscript{1}H-NMR spectrum of viscosity-sensitive monomer, MECVJ.}
    \label{NMRMECVJ}
\end{figure}

\subsection{Synthesis of the viscosity-sensitive polymer poly(DMA-\textit{s}-MECVJ) P2}

\begin{figure}[!htb]
\centering
\includegraphics[width=0.5\textwidth]{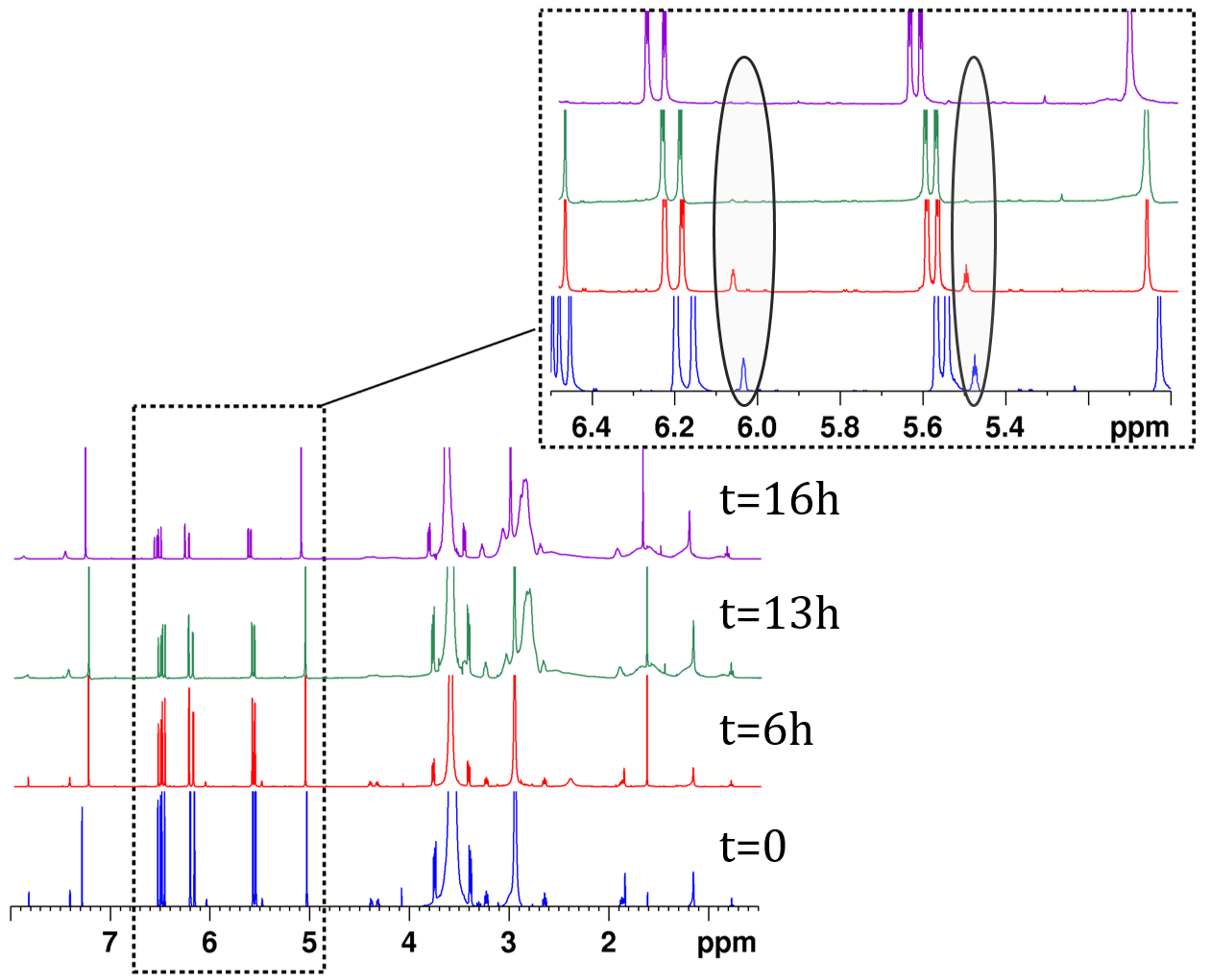}
\caption{\corr{$^1$1H-NMR monitoring of the synthesis of  poly(DMA-\textit{s}-MECVJ), P2, in \ce{CDCl3}. The inset shows the disappearance of peaks at 5.5 and 6.1 ppm corresponding to the vinylic protons of the MECVJ monomer.}}
\label{fig:polymer-NMR-monitoring}
\end{figure}

\corr{\textsuperscript{1}H-NMR spectra of the synthesized viscosity-sensitive fluorescent polymer P2 and monitoring over time. The evolution of MECVJ and DMA polymerization for P2 is shown in Figure~\ref{fig:polymer-NMR-monitoring}. MECVJ can be seen to be fully incorporated into the copolymer before full conversion of DMA is reached, with the disappearance of the peaks at 5.5 and 6.1 ppm which correspond to the vinylic protons of MECVJ monomer.}

\subsection{Synthesis of the viscosity-sensitive polymer poly(DMA-\textit{s}-MECVJ) P3}

\begin{figure}[!htb]
\centering
\includegraphics[width=0.5\textwidth]{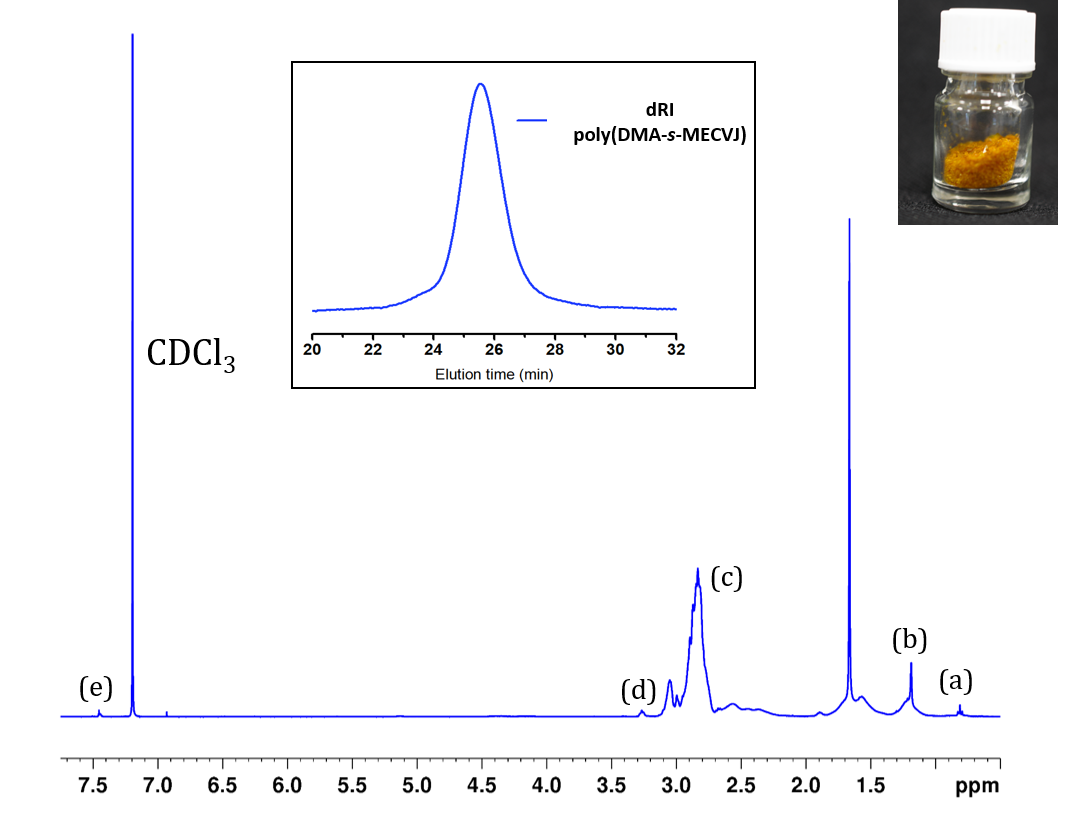}
\caption{\corr{$^1\text{H-NMR}$ spectra of poly(DMA-\textit{s}-MECVJ) P3. (a), (b), and (d) represent the characteristic proton peaks of the RAFT agent, (c) the polymerized DMA, and (e) the polymerized MECVJ. The inset chromatogram obtained via SEC in DMF represents the differential refractive signal of the copolymer. Top-right inset: an image of P3 synthesized in this work.}}
\label{fig:NMR_polymer}
\end{figure}

\corr{\textsuperscript{1}H-NMR spectra of the synthesized viscosity-sensitive fluorescent polymer P3 and SEC characterization. Results of the analysis of quality of the polymer P3 is displayed in Figure~\ref{fig:NMR_polymer}.}

\subsection{Response to viscosity of poly(DMA-\textit{s}-MECVJ) P1, P3 and P4}

\corr{Emission spectra and calibration curves of polymers P1, P3, and P4, results for P2 being displayed in main text. Table specifies wavelength of maximum absorption and emission for the different polymers.}

\begin{table}[!h]
\begin{center}
\begin{tabular}{|c| c| c|} 

 \hline
  \textbf{Polymer} &\textbf{$\lambda_{\textbf{absorption}}$(nm)} &\textbf{$\lambda_{\textbf{emission}}$(nm)}\\ 
 \hline
 P1  &  465 & 504 \\ 
 \hline
 P2  & 461 & 510 \\
 \hline
 P3  & 461 & 506 \\ 
 \hline
P4  & 458 & 506 \\
 \hline

\end{tabular}
\caption{Absorption and emission maxima \corr{in glycerol} of different poly(DMA-MECVJ) synthesized in this work. }
\label{table:summary-polymer-wavelengths}
\end{center}
\end{table}

\begin{figure}[!htb]
\centering
\includegraphics[width=0.5\textwidth]{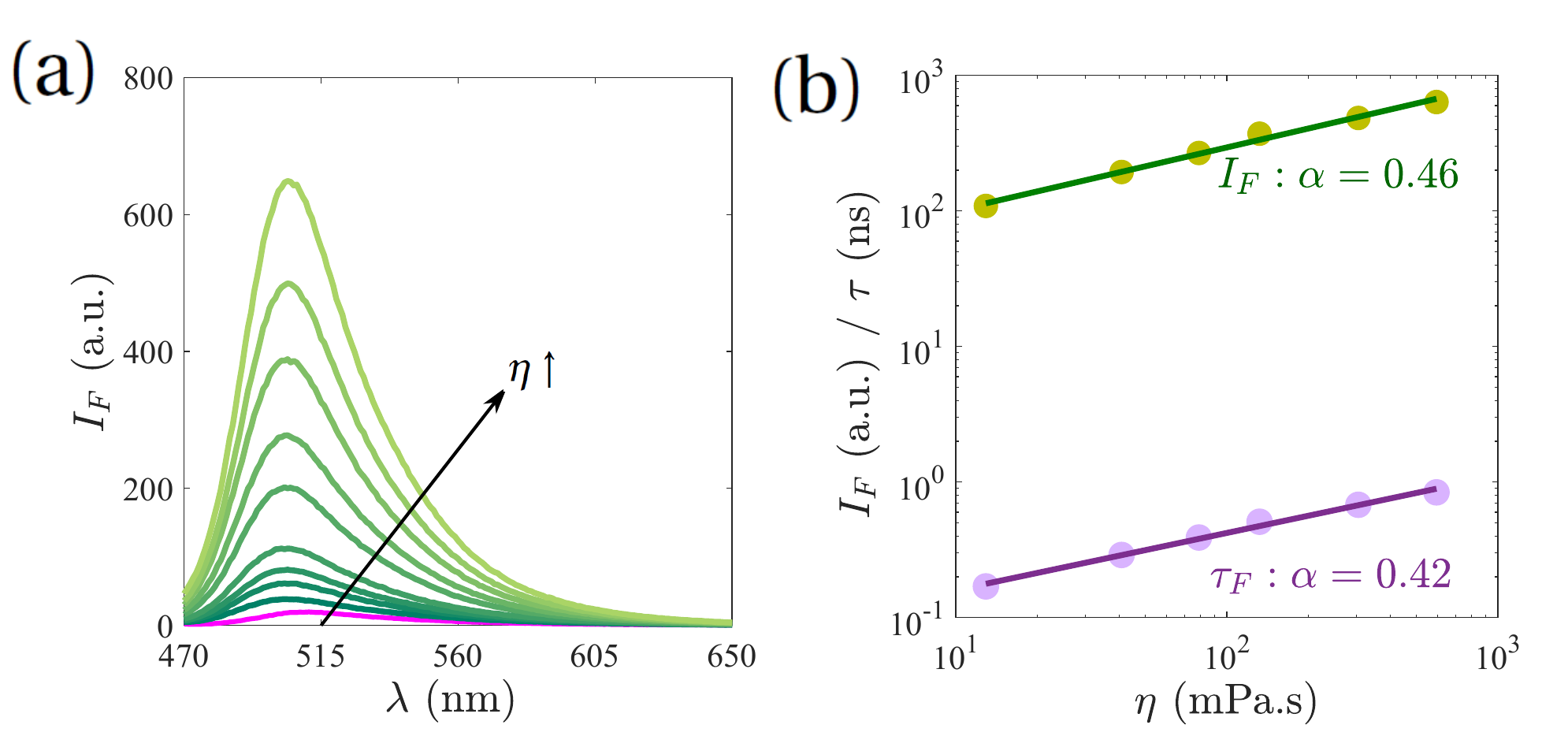}
\caption{\corr{(a) Emission spectra of P1 in different glycerol/DMSO mixtures of respective viscosity $\eta = 1$, $2$, $4$, $7$, $12$, $40$, $78$, $132$, $305$ and $\SI{593}{\milli\pascal\second}$ with excitation wavelength $\SI{450}{\nano\meter}$. (b) Corresponding calibration curves in logarithmic scale of P1. Intensity $I_\text{F}$ is represented with green points and lifetime $\tau_\text{F}$ with purple points.}}
\label{fig:carac_P1}
\end{figure}

\begin{figure}[!htb]
\centering
\includegraphics[width=0.5\textwidth]{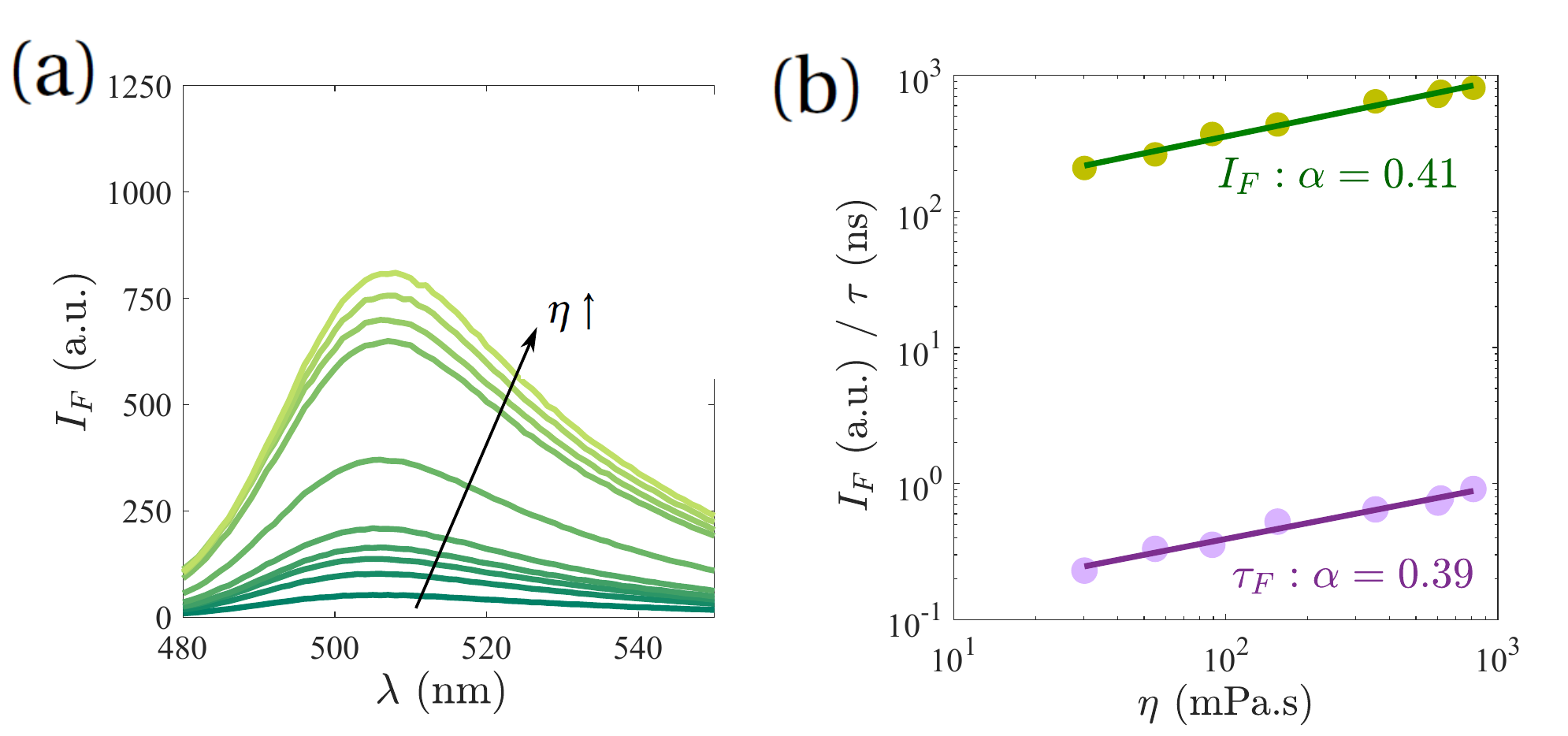}
\caption{\corr{(a) Emission spectra of P3 in different glycerol/DMSO mixtures of respective viscosity $\eta = 2$, $8$, $15$, $18$, $30$, $89$, $358$, $606$, $623$ and $\SI{818}{\milli\pascal\second}$ with excitation wavelength $\SI{450}{\nano\meter}$. (b) Corresponding calibration curves in logarithmic scale of P3. Intensity $I_\text{F}$ is represented with green points and lifetime $\tau_\text{F}$ with purple points.}}
\label{fig:carac_P3}
\end{figure}

\begin{figure}[!htb]
\centering
\includegraphics[width=0.5\textwidth]{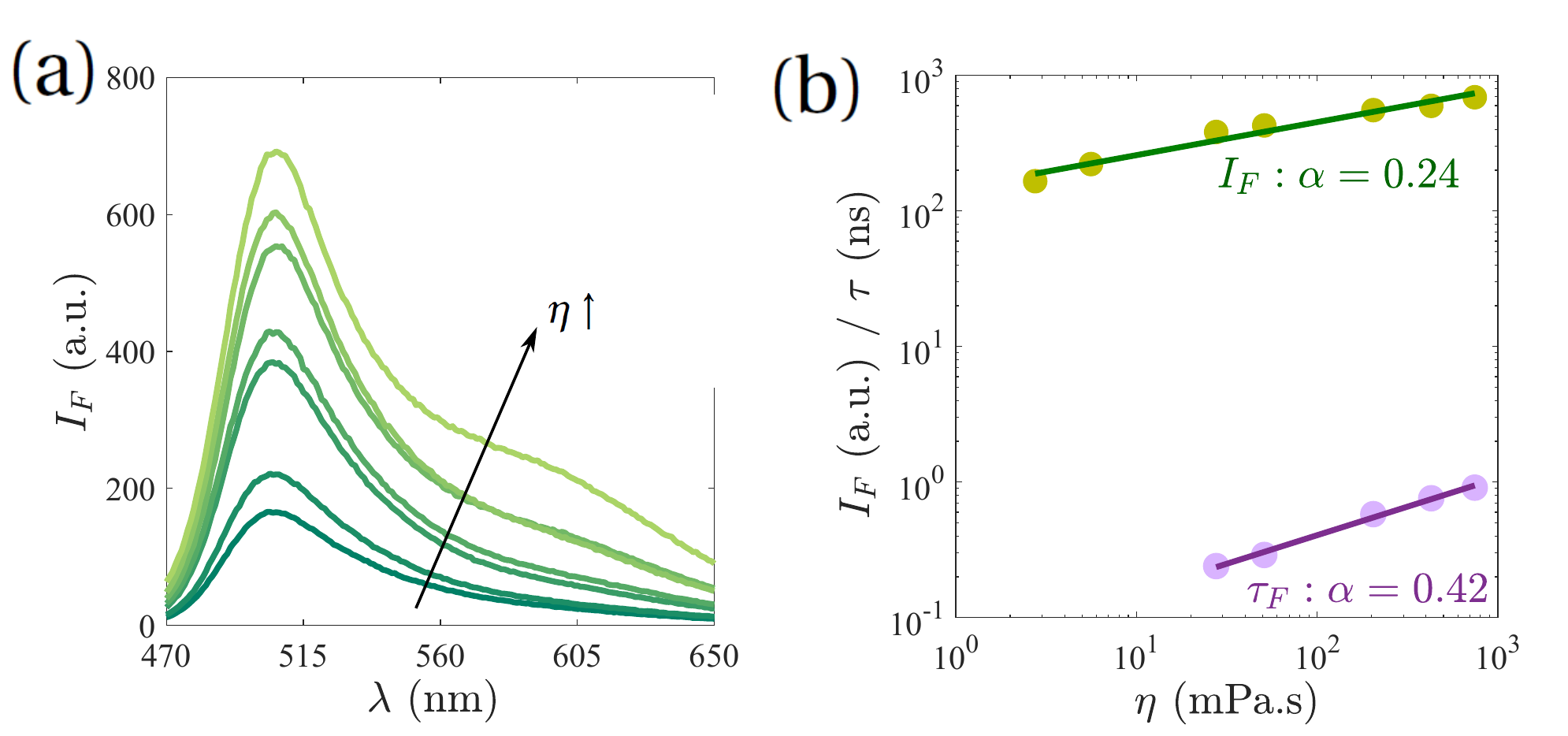}
\caption{\corr{(a) Emission spectra of P4 in different glycerol/DMSO mixtures of respective viscosity $\eta = 3$, $6$, $28$, $51$, $204$, $428$ and $\SI{745}{\milli\pascal\second}$ with excitation wavelength $\SI{450}{\nano\meter}$. (b) Corresponding calibration curves in logarithmic scale of P4. Intensity $I_\text{F}$ is represented with green points and lifetime $\tau_\text{F}$ with purple points.}}
\label{fig:carac_P4}
\end{figure}

\subsection{XPS analysis of grafted poly(DMA-\textit{s}-MECVJ)}

\corr{XPS spectra of bare glass-slide and grafted poly(DMA-\textit{s}-MECVJ) on a gold-coated glass slide.}

\begin{figure*}[!htbp]
\centering
\begin{subfigure}{0.8\textwidth}
    \caption{}
    \includegraphics[width=\textwidth]{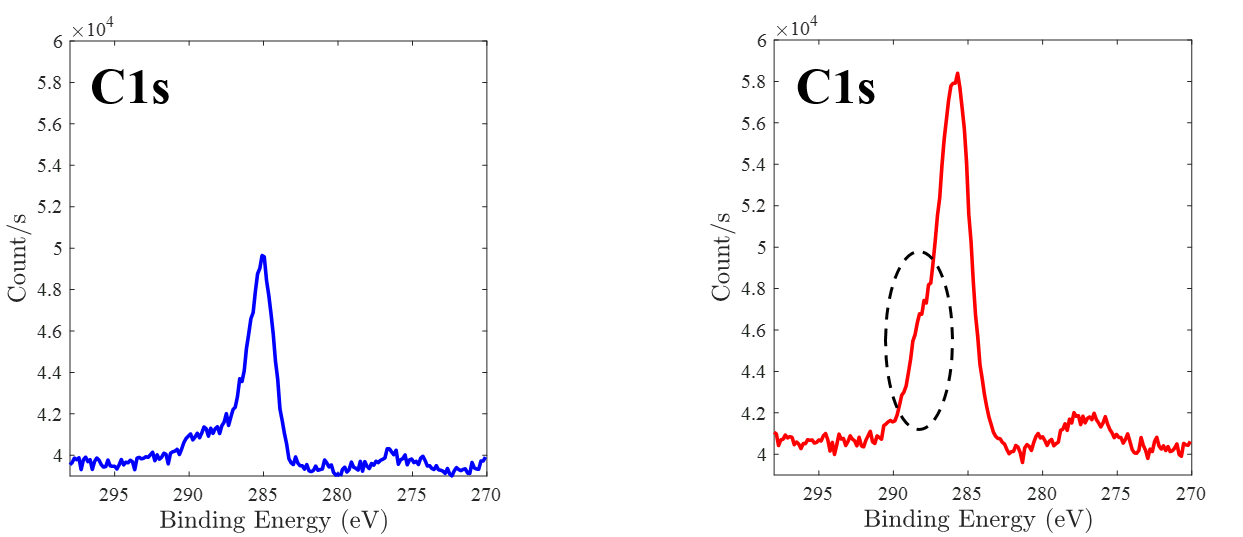}
    \label{fig:XPS-C1s}
    \end{subfigure}
\begin{subfigure}{0.8\textwidth}
    \caption{}
    \includegraphics[width=\textwidth]{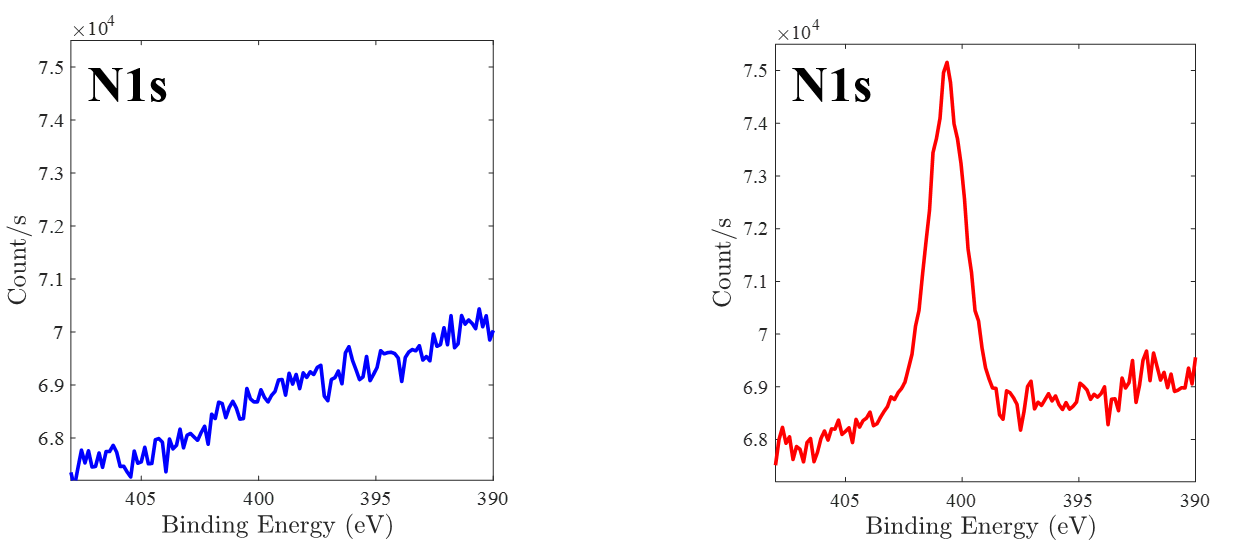}
    \label{fig:XPS-N1s}
    \end{subfigure}
\begin{subfigure}{0.8\textwidth}
    \caption{}
    \includegraphics[width=\textwidth]{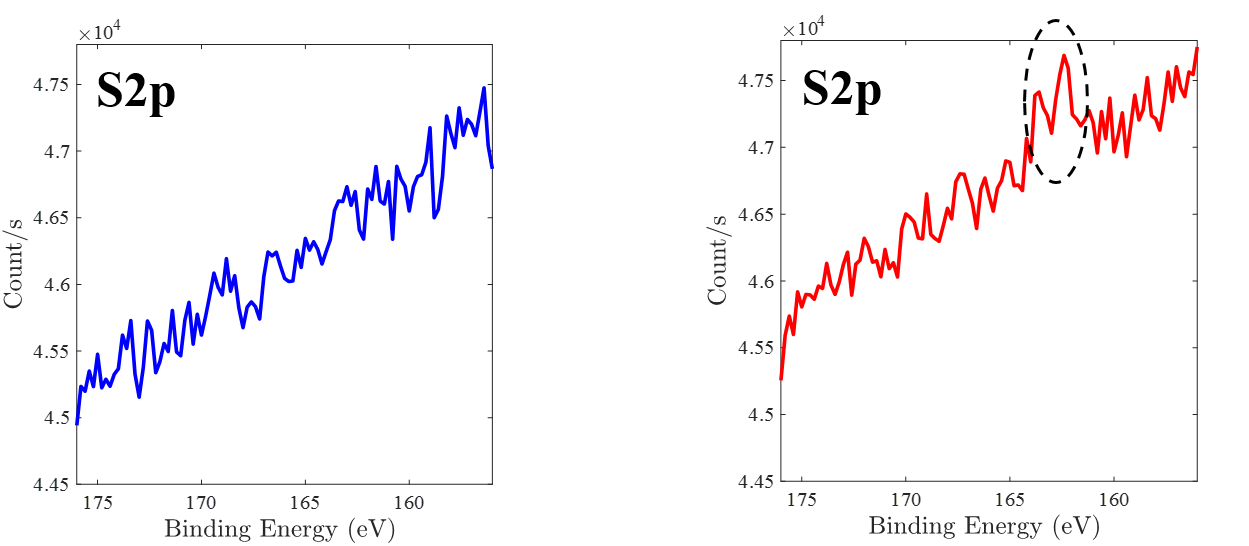}
    \label{fig:XPS-S2P}
    \end{subfigure}
    \caption{XPS analysis of a non-grafted gold glass slide (in blue) and of grafted poly(DMA-\textit{s}-MECVJ) on a gold-coated glass slide (in red). (a) Representative C 1s spectrum. (b) Representative N 1s spectrum. (c) Representative S 2p spectrum. The significance of the dashed lines is explained in the accompanying text.}
    \label{fig:XPS-results}
\end{figure*}

\end{document}